\documentclass[conference]{IEEEtran}
\IEEEoverridecommandlockouts
\usepackage[english]{babel}
\usepackage{lipsum}
\usepackage{indentfirst}
\usepackage{algorithm}
\usepackage{mathtools}
\usepackage{balance}
\usepackage{commath}
\usepackage{comment}
\usepackage{dblfloatfix}
\usepackage{amsmath}
\usepackage{multirow}
\usepackage{amssymb}
\usepackage{algpseudocode}
\usepackage{float}
\usepackage{graphicx}
\usepackage{subcaption}
\usepackage{bm}
\usepackage[font={small}]{caption}
\usepackage{array}
\usepackage{verbatim}
\usepackage{xcolor}
\usepackage[noadjust]{cite}
\def\bibTeX{{\rm b\kern-.05em{\sc i\kern-.025em b}\kern-.08em
    T\kern-.1667em\lower.7ex\hbox{E}\kern-.125emX}}
\begin{document}

\title{Reliability Analysis of Fault Tolerant Memory Systems\\}

\author{
\IEEEauthorblockN{
Yagmur Yigit\IEEEauthorrefmark{1}, 
Leandros Maglaras\IEEEauthorrefmark{1}, 
Mohamed Amine Ferrag\IEEEauthorrefmark{2}, 
Naghmeh Moradpoor\IEEEauthorrefmark{1}, Georgios Lambropoulos\IEEEauthorrefmark{3} }
\IEEEauthorblockA{
\IEEEauthorrefmark{1} School of Computing, Engineering and The Build Environment, Edinburgh Napier University, UK \\
\IEEEauthorrefmark{2} Technology Innovation Institute, United Arab Emirates, UAE \\
\IEEEauthorrefmark{3} University of Piraeus, Department of Informatics, Piraeus Greece \\
Email:  yagmur.yigit@napier.ac.uk, L.Maglaras@napier.ac.uk, \\ mohamed.ferrag@tii.ae, N.Moradpoor@napier.ac.uk, ; george@lambropoulos.com}
}


\maketitle

\IEEEpubidadjcol

\begin{abstract}
This paper delves into a comprehensive analysis of fault-tolerant memory systems, focusing on recovery techniques modelled using Markov chains to address transient errors. The study revolves around the application of scrubbing methods in conjunction with Single Error Correction and Double Error Detection (SEC-DED) codes. It explores three primary models: 1) Exponentially distributed scrubbing, involving periodic checks of memory words within exponentially distributed time intervals; 2) Deterministic scrubbing, featuring regular, periodic word checks; and 3) Mixed scrubbing, which combines both probabilistic and deterministic scrubbing approaches. The research encompasses the estimation of reliability and Mean Time to Failure (MTTF) values for each model. Notably, the findings highlight the superior performance of mixed scrubbing over simpler scrubbing methods in terms of reliability and MTTF.
\end{abstract}


\section{Introduction}
When designing a fault-tolerant memory system, one encounters a plethora of choices. A robust fault-tolerant system should possess the capability to detect, diagnose, isolate, mask, compensate for, and recover from errors and faults. In the models examined in this article, SEC-DED (Single Error Correction-Double Error Detection) codes are employed \cite{samanta2021compact}, as they strike a balance between minimal overhead (attributed to extra parity bits) and highly efficient error detection. Additionally, memory scrubbing proves to be a highly effective approach for recovering from transient faults caused by environmental disruptions and intermittent faults arising from inherent weaknesses within the circuit \cite{vlagkoulis2022configuration}, \cite{2016EAI}. It is a technique that reads words, checks their correctness, and rewrites the corrected data in its initial position \cite{kwon2020reliability}, \cite{maglaras2022reliability}. 
A previous study analyzed the reliability of scrubbing recovery techniques for memory systems, emphasizing the use of SEC-DED codes \cite{IEEE90}. This research examined two models: exponentially distributed scrubbing and deterministic scrubbing. The authors derived reliability and mean-time-to-failure (MTTF) equations and compared the results with memory systems lacking redundancy and utilizing only SEC-DED codes.
Probabilistic scrubbing is based on the fact that every time the program in execution processes a word, it is read and checked for correctness. This technique offers a good solution for environments where all words are addressed at the same rate by the program \cite{baraza2020proposal}. But in real systems, it is almost certain that some words are more often used than others, making this technique inadequate \cite{Bridge2023}. Deterministic scrubbing, on the other hand, depends on a mechanism that reads every word and checks for its correctness periodically, thus improving system reliability and MTTF \cite{scargall2020reliability}, \cite{maglaras2022combining}. This technique achieves better reliability than the probabilistic one but demands more complex architecture, especially when a small scrubbing interval is required. Mixed scrubbing is finally based on the fact that the system uses exponentially distributed scrubbing with the addition of a mechanism that reads every word. This model achieves better reliability than the first one and achieves better reliability from the second one with a much simpler architecture (scrubbing interval required). 
The main contribution of this paper is that it thoroughly analyses the three techniques and computes and compares the reliability and MTTF of RAM of different word bit sizes.

The rest of this paper is organized as follows: Section 2 presents the terminology, assumptions, and notation used in this article. Section 3 presents the reliability analysis of the memory system. Section 4 presents the experimental results, and Section 5 concludes the article.

\section{Terminology, Notations, Assumptions}

\begin{table}[htbp]
\caption{Notations} 
\label{tab:Notation}
\centering
\begin{tabular}{ll}
Notation            & Definition       \\\hline\hline
$\lambda$           & Failure Rate per Memory Bit \\
$w$                  & Number of Bits per Word    \\
$c$                  & Number of Check Bits per Word     \\
$L$                 & Error Rate of One Word     \\
$M$                 & Memory Size in Number of Words   \\
$r(t)$              & Reliability of One Word\\
$R(t)$            & Reliability of the Memory System  \\
$T$               & Interval of Deterministic Scrubbing \\
$n$               & Number of Intervals of Deterministic Scrubbing \\
$\mu$               & Rate of Probabilistic Scrubbing \\
\textit{MB}               & Megabyte  \\
\textit{SEC-DED }              & \begin{tabular}[c]{@{}l@{}}Single Error Correction,\\Double Error Detection Process in Memory Word\end{tabular}\\
\textit{Memory System }              & \begin{tabular}[c]{@{}l@{}}Computer Hardware Associated with\\ Data Storage and Retrieval\end{tabular}  \\
\textit{Memory Word }              & Group of Bits Capable of Storing Data  \\
$MTTF$               & Mean Time to Failure  \\
\hline
\end{tabular}
\end{table}

The notations are presented in Table~\ref{tab:Notation}, and our underlying assumptions can be summarized as follows:
\begin{enumerate}
 \item Transient faults manifest with a Poisson distribution.
 \item Failures of individual bits are statistically uncorrelated.
 \item The control and correction mechanisms integrated within the memory operate flawlessly.
 \item Each word is treated as a statistically independent entity.
 \item It is possible for an erroneous bit to recover due to the presence of another error.
 \end{enumerate}

\section{Reliability Analysis of Memory Systems}
In this section, we present a reliability analysis for memory systems that use deterministic, probabilistic, or mixed scrubbing techniques.

\subsection{Exponentially Distributed Scrubbing}

The scrubbing mechanism employed in this method relies on the premise that every time a memory location is accessed, it undergoes a check and correction process if necessary. The system utilizes a Single Error Correction-Double Error Detection (SEC-DED) code to detect and rectify errors. However, it's important to note that this code may fail when a memory word contains a sufficient number of multiple-bit errors \cite{2023GCWshp}. 
We represent with S0, S1, and S2 the states of having zero, one or two memory errors in an SEC-DEC-protected memory word, respectively.
The interval between two consecutive accesses to the same memory location follows an exponential distribution with a rate denoted as $\mu$ \cite{YY2022}. While this assumption may not hold true for all memory systems, the operating system can enforce the use of exponential scrubbing.

\begin{figure}[htbp]
\centering
\includegraphics[width=3in]{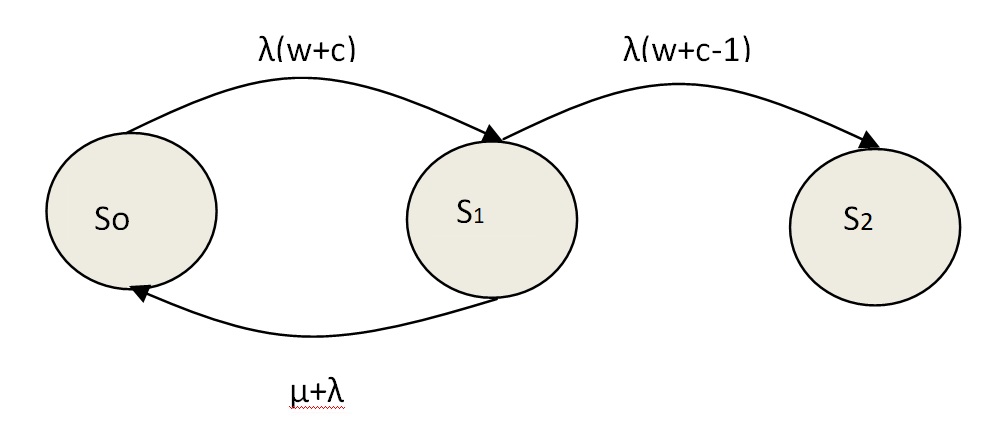}
\caption{State transition rate diagram for probabilistic scrubbing}
\label{fig1}
\end{figure}

A Markov chain is employed to model the recovery scenario, which features as many states as the error correction code being used \cite{2023Nature}. Fig.~\ref{fig1} depicts the Markov chain representing this model.

Assuming that the error rate of a memory word with w data and c check bits is $\lambda$(w + c), assuming all bits in the word fail statistically independently with a failure rate $\lambda$. The error rate of a memory word when being in state a is:

\begin{equation}
    L = \lambda + \lambda +\lambda +\lambda + .... + \lambda  = \lambda (w+c)
\end{equation}

The transition from state $S_0$ to state $S_1$ occurs when:

\begin{itemize}
    \item The scrubbing mechanism is activated (the word is addressed by the program in execution) [rate $\mu$].
    \item An error occurs in the faulty bit cancelling the initial one [rate $\lambda$].
\end{itemize}

When the word is in state $S_1$, the occurrence of an error in any bit other than the faulty leads the memory word to state $S_2$ [rate $\lambda(w+c-1)$].

The differential equations that describe the system are:

\begin{equation}
dP_0(t)/dt = -\lambda(w+c)P_0(t) + (\lambda+\mu)P_1(t)
\end{equation}
\begin{equation}
dP_1(t)/dt = \lambda(w+c)P_0(t) - (\lambda(w+c) +\mu)P_1(t)
\end{equation}
\begin{equation}
dP_2(t)/dt = (\lambda(w+c-1))P_1(t)
\end{equation}

The solution to the aforementioned equations can be derived by taking their Laplace transforms into account \cite{2019DL_M} while also considering their respective initial conditions:
$$ P_0(0) =1, P_1(0) = 0, P_2(0)=0$$

The final form of the reliability is:
\begin{equation}
r(t) = \frac{a_{1p}}{a_{1p} - a_{2p}}exp(-a_{2p}t) + \frac{a_{2p}}{a_{2p} - a_{1p}}exp(-a_{1p}t)
\end{equation}


The reliability of the memory system with M words is:

\begin{equation}
R(t) = r(t)^M
\end{equation}

The MTTF is:
\begin{equation}
MTTF = \int_{0}^{\infty} R(t) \,dt
\end{equation}

The final form of MTTF is:
\begin{equation}
MTTF = \frac{4(\mu +\lambda)}{M(\lambda(2w+2c-1))^2}
\end{equation}

\subsection{Deterministic Scrubbing}
The deterministic scrubbing used in this model is based on the fact that one or more processors cycle through the memory system, scrub the memory words, and correct any transient errors \cite{2020Resiliency}. The cycling time is T, and checking the word is executed in parallel with the normal operations when this part of the memory is idle. The memory system fails every time that two errors co-exist in the same memory word. The time interval T can be reduced depending on the hardware mechanism that is used.

Deterministic scrubbing ensures that all the memory words are checked and corrected, if possible, at the same rate, improving the reliability and MTTF of the memory system. The complexity of the hardware, on the other hand, is a disadvantage of this technique that must be used only in very noisy environments when transient errors occur at a high rate \cite{Reslient2017}. 

\begin{figure}[htbp]
\centering
\includegraphics[width=3in]{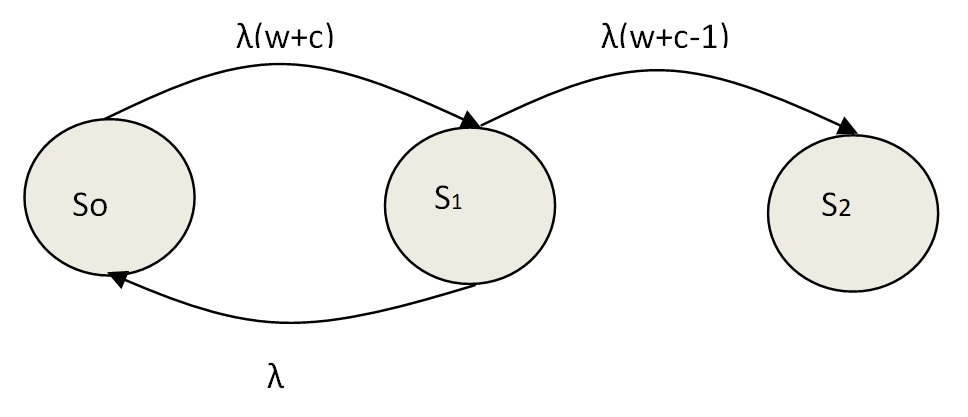}
\caption{State transition rate diagram for deterministic scrubbing}
\label{fig2}
\end{figure}

The behaviour of a memory word in the time interval $0<t<T$ can be modelled using the Markov chain, as can be seen in Fig.~\ref{fig2}.

Assuming that the error rate of every bit is $\lambda$ and that the different bits are statistically independent, the error rate of a word when being in state a is:

\begin{equation}
    L = \lambda + \lambda +\lambda +\lambda + .... + \lambda  = \lambda (w+c)
\end{equation}

The transition from state $S_1$ to state $S_0$ occurs when an error occurs in the faulty bit, cancelling the initial one [rate $\lambda$].

When the word is in state $S_1$, the occurrence of an error in any bit other than the faulty leads the memory word to state $S_2$ [rate $\lambda(w+c-1)$].

The differential equations that describe the system are:

\begin{equation}
dP_0(t)/dt = -\lambda(w+c)P_0(t) + (\lambda)P_1(t)
\end{equation}
\begin{equation}
dP_1(t)/dt = \lambda(w+c)P_0(t) - (\lambda(w+c))P_1(t)
\end{equation}
\begin{equation}
dP_2(t)/dt = (\lambda(w+c-1))P_1(t)
\end{equation}

By using the Laplace transforms of the aforementioned equations and keeping in mind that the following initial conditions apply:
$$ P_0(0) =1, P_1(0) = 0, P_2(0)=0$$

The reliability of one word in the time interval (0,t) is given by the type:
\begin{equation}
r_0(t) = \frac{a_{1d}}{a_{1d} - a_{2d}}exp(-a_{2d}t) + \frac{a_{2d}}{a_{2d} - a_{1d}}exp(-a_{1d}t)
\end{equation}

The reliability of a word at a specific moment is established by the multiplication of two probabilities: the probability of its survival throughout all the preceding scrubbing intervals and the probability of enduring the additional time until the present interval.

\begin{equation}
r(t) = [r_0(t)]^nr_o(x),t-nT+x,0\leq x\leq T
\end{equation}

At time $nT$, the reliability of a word is determined by the probability that it has endured without more than one error co-existing in the word during all the preceding n time intervals. Consequently, it can be expressed as the reliability at time $T$ raised to the power of n. Since all memory words are considered to be statistically independent, the overall reliability of the memory system can be calculated as follows:

\begin{equation}
R(t) = r(t)^M
\end{equation}

The MTTF is:
\begin{equation}
MTTF = \int_{0}^{\infty} R(t) \,dt
\end{equation}

\begin{figure}[htbp]
\centering
\includegraphics[width=3.3in]{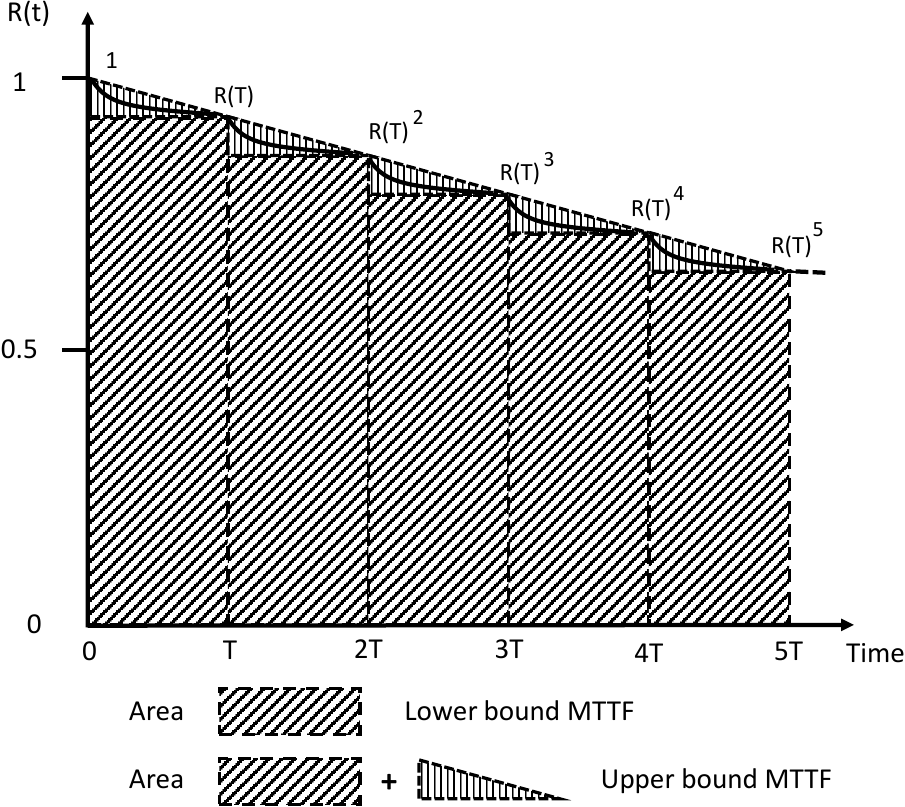}
\caption{MTTF for deterministic scrubbing}
\label{fig3}
\end{figure}

The calculation of the MTTF can be achieved using the following approximation. Fig.~\ref{fig3} shows that the MTTF has an upper (MTTFu) and lower bound (MTTFl). Using that, we can say that $MTTFl<MTTF<MTTFu$.

\begin{equation}
MTTFl= \frac{TR(T)}{1-R(T)}
\end{equation}

\begin{equation}
MTTFu= \frac{T[1+R(T)]}{2[1-R(T)]}
\end{equation}

So the MTTF bounds are:

\begin{equation}
\frac{TR(T)}{1-R(T)}< MTTF < \frac{T[1+R(T)]}{2[1-R(T)]}
\end{equation}

\begin{table*}[t]
\caption{Summary of Three Systems} 
\label{tab:summary}
\centering
\begin{tabular}{ccc}
Memory System  & Reliability R(t)& MTTF\\
\hline\hline
Probabilistic Scrubbing& $[\frac{a_{1p}}{a_{1p} - a_{2p}}exp(-a_{2p}t) + \frac{a_{2p}}{a_{2p} - a_{1p}}exp(-a_{1p}t)]^M$  & $\frac{4\mu}{M(2\lambda+2w-1)^2}$\\  \\ 
Deterministic Scrubbing& $[R_d(T)]^n R_d(x), t=nT+x, 0\leq x\leq T$& $\frac{T[1+R_d(T)]}{2[1-R_d(T)]}$\\ \\ 
Mixed Scrubbing& $[R_m(T)]^n R_m(x), t=nT+x,0 \leq x\leq T$ & $\frac{T[1+R_m(T)]}{2[1-R_m(T)]}$\\ 

\hline
\end{tabular}
\end{table*}

\subsection{Mixed Scrubbing}
The model described in this section combines probabilistic and deterministic scrubbing. Whenever a word is addressed, it is checked for correctness and rewritten in its initial condition. The rate of probabilistic scrubbing is $\mu$. In order to enhance the system's reliability and the MTTF, a specialized hardware mechanism is introduced to systematically traverse through the memory and scrub each word \cite{maglaras2022mean}. The time interval of the deterministic scrubbing is $T$.

The Markov chain that describes the behaviour of one memory word in the time interval (0, t) is the same as the probabilistic scrubbing \cite{2023MDPI}. The transition of the system from one state to the other follows the same rules as in the probabilistic scrubbing, and the final form of the reliability of one memory word in the time interval (0, t) is:
\begin{equation}
r_0(t) = \frac{a_{2m}}{a_{2m} - a_{1m}}exp(-a_{1m}t) + \frac{a_{1m}}{a_{1m} - a_{2m}}exp(-a_{2m}t)
\end{equation}

Using the same reasoning as in the deterministic scrubbing method, we can find  that the final form of the reliability of the memory word system at time t is:

\begin{equation}
R(t) = r(t)^M
\end{equation}

Where: 

\begin{equation}
r(t) = [r_0(t)]^nr_o(x),t-nT+x,0\leq x\leq T
\end{equation}

MTTF is also calculated in a similar way as in the deterministic method: 

\begin{equation}
\frac{TR(T)}{1-R(T)}< MTTF < \frac{T[1+R(T)]}{2[1-R(T)]}
\end{equation}

The summary of the reliability and MTTF can be seen in Table~\ref{tab:summary} for the three systems described above. For deterministic and mixed scrubbing, the upper limit of MTTF (MTTFu) is used.

\section{Evaluation}

\begin{table}[htbp]
\caption{MTTF of memory systems with 32-bit words \\ $\lambda= 0,00001 upsets/bit/day$, w=32, c=7, T=10 sec, $\mu =0,1 sec$} 
\label{tab1}
\centering
\begin{tabular}{cccc}
Memory size &Probabilistic& Deterministic& Mixed\\
\hline\hline
1 MB& 3,6E+05& 6,35E+05& 1,02E+06\\  
2 MB& 1,8E+05& 3,18E+05& 5,08E+05\\ 
4 MB& 9,0E+04& 1,58E+05& 2,54E+05\\ 
8 MB& 4,5E+04& 7,945E+04& 1,27E+05\\ 
16 MB& 2,2E+04& 3,97E+04& 6,35E+04\\ 
32 MB& 1,12E+04& 1,99E+04& 3,18E+04\\ 
64 MB& 5,62E+03& 9,93E+03& 1,59E+04\\ 
128 MB& 2,81E+03& 4,96E+03& 7,94E+03\\ 
\hline
\end{tabular}
\end{table}
\begin{table}[htbp]
\caption{MTTF of Memory Systems with 64-bit Words \\ \textit{$\lambda= 0,00001 upsets/bit/day$, w=64, c=8, T=10 sec, $\mu =0,1 sec$}} 
\label{tab2}
\centering
\begin{tabular}{cccc}
Memory size &Probabilistic& Deterministic& Mixed\\
\hline\hline
1 MB& 1,66E+05& 2,61E+05& 4,47E+06\\  
2 MB& 8E+04& 1,30E+05& 2,23E+05\\ 
4 MB& 4,15E+04& 6,52E+04& 1,12E+05\\ 
8 MB& 2,08E+04& 3,26E+04& 5,58E+04\\ 
16 MB& 1,04E+04& 1,63E+04& 2,79E+04\\ 
32 MB& 5,19E+03& 8,14E+03& 1,40E+04\\ 
64 MB& 2,59E+03& 4,07E+03& 6,98E+03\\ 
128 MB& 1,30E+03& 2,04E+03& 3,19E+03\\ 
\hline
\end{tabular}
\end{table}
\begin{table}[htbp]
\caption{MTTF of Memory Systems with 32-bit Wwords \\ $\lambda= 0,0001 upsets/bit/day$, w=32, c=7, T=10 sec, $\mu =0,1 sec$} 
\label{tab3}
\centering
\begin{tabular}{cccc}
Memory size &Probabilistic& Deterministic& Mixed\\
\hline\hline
1 MB& 3,6E+03& 7,22E+03& 9,77E+03\\  
2 MB& 1,8E+03& 3,61E+03& 4,89E+03\\ 
4 MB& 9E+02& 1,80E+03& 2,44E+03\\ 
8 MB& 4,5E+02& 9,02E+02& 1,22E+03\\ 
16 MB& 2,25E+02& 4,51E+02& 6,11E+02\\ 
32 MB& 1,12E+02& 2,26E+02& 3,06E+02\\ 
64 MB& 5,62E+01& 1,13E+02& 1,53E+02\\ 
128 MB& 2,61E+01& 5,64E+01& 7,64E+01\\ 
\hline
\end{tabular}
\end{table}
\begin{table}[htbp]
\caption{MTTF of memory systems with 64-bit words \\ $\lambda= 0,0001 upsets/bit/day$, w=34, c=8, T=10 sec, $\mu =0,1 sec$} 
\label{tab4}
\centering
\begin{tabular}{cccc}
Memory size &Probabilistic& Deterministic& Mixed\\
\hline\hline
1 MB& 1,66E+03& 3,32E+03& 4,52E+03\\  
2 MB& 8,30E+02& 1,66E+03& 2,26E+03\\ 
4 MB& 4,15E+02& 8,29E+02& 1,13E+03\\ 
8 MB& 2,08E+02& 4,14E+02& 5,64E+02\\ 
16 MB& 1,045E+02& 2,07E+02& 2,82E+02\\ 
32 MB& 5,19E+01& 1,03E+02& 1,41E+02\\ 
64 MB& 2,59E+01& 5,18E+01& 7,05E+01\\ 
128 MB& 1,30E+01& 2,59E+01& 3,53E+01\\ 
\hline
\end{tabular}
\end{table}

We conducted extensive simulated scenarios in order to evaluate the MTTF and reliability values for different memory systems with various error rates and memory sizes. The value $\lambda$ represents the error rate per bit per day, and the values $1/\mu$, T are scrubbing intervals in seconds.

\begin{table}[htbp]
\caption{The Comparison of MTTF of the Three Memory Systems} 
\label{tab5}
\centering
\begin{tabular}{cc}
\begin{tabular}[c]{@{}l@{}}(T=10, $\mu$=0.1),\\ $1  < M < 128 $\end{tabular}  & Ratio ( $MTTF_a$/$MTTF_b$)\\
\hline\hline
Deterministic / Probabilistic & 1.6 - 2\\  
Mixed / Deterministic & 1.6 - 1.75\\ 
Mixed / Probabilistic & 2.5 - 3.5\\ 
\hline
\end{tabular}
\end{table}

\begin{table*}[htbp]
\caption{The Effect of Parameters T and $\mu$ to the Values of MTTF of the Memory Systems \textit{[$\lambda$=0.00001, (w+c)=72, M=128MB]}} 
\label{tab6}
\centering
\begin{tabular}{cccc}
\textbf{(a)}  & $\mu$=0.1 & $\mu$=0.01 & $\mu$=0.001 \\
\hline\hline
Probabilistic & 1300 & 130 & 13\\  
Mixed  & 3490 & 2618 & 2528 \\\hline
\\

\textbf{(b)}  & T=10 & T=100 & T=1000 \\
\hline\hline
Deterministic & 2400 & 260 & 26\\  
Mixed  & 3490 & 1434 & 1300 \\\hline \\

\textbf{(c)} & T=10, $\mu$=0.1 & T=100, $\mu$=0.01 & T=1000, $\mu$=0.001\\
\hline\hline
Probabilistic & 1300 & 130 & 13\\  
Deterministic & 2400 & 260 & 26\\  
Mixed  & 3490 & 352 & 35\\
\hline
\end{tabular}
\end{table*}

\begin{figure}[htbp]
\centering
\includegraphics[width=3.2in]{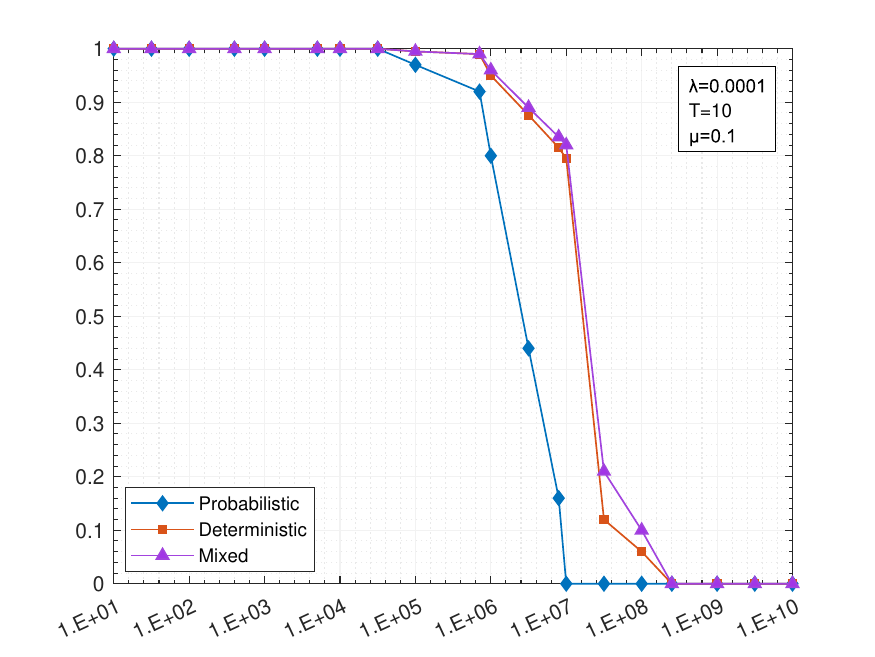}
\caption{Reliability of memory systems of 128 MB of 32-bit words for various memory systems.}
\label{fig:4.3}
\end{figure}
\begin{figure}[htbp]
\centering
\includegraphics[width=3.2in]{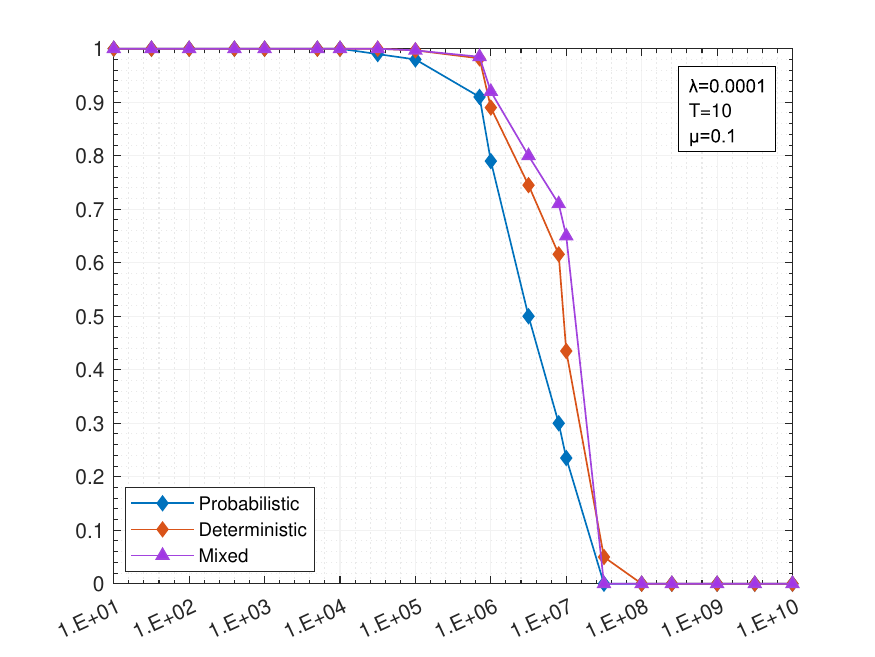}
\caption{Reliability of memory systems of 128 MB of 64-bit words for various memory systems.}
\label{fig:4.4}
\end{figure}
\begin{figure}[htbp]
\centering
\includegraphics[width=3.2in]{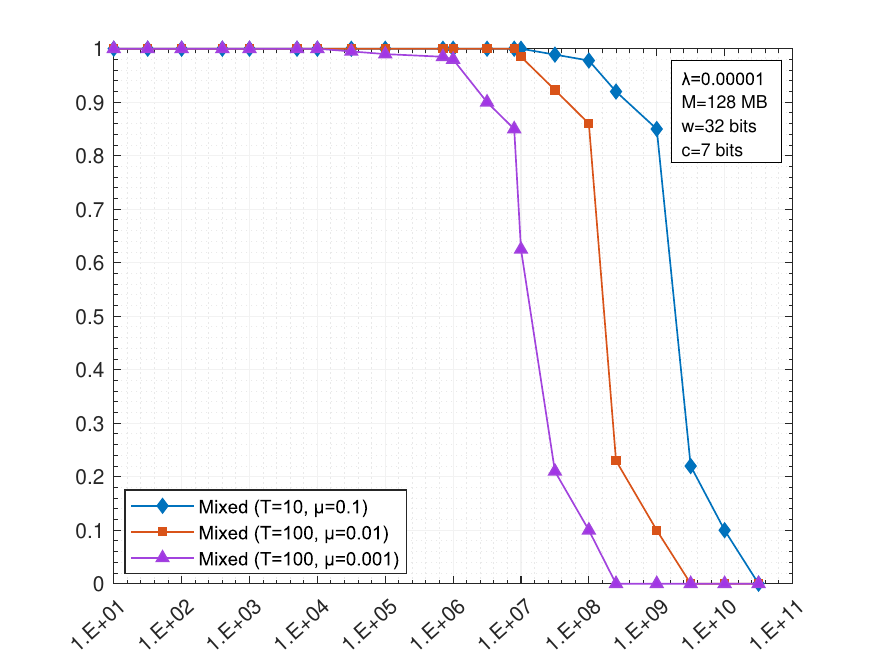}
\caption{Reliability of memory systems with mixed scrubbing for different values of parameters T and $\mu$.}
\label{fig:4.7}
\end{figure}

When we compare the values provided in Table~\ref{tab1}, Table~\ref{tab2}, Table~\ref{tab3}, and Table~\ref{tab4}, we observe that the ratio of the MTTF for deterministic scrubbing to the MTTF for probabilistic scrubbing falls in the range of 1.6 to 2. This comparison pertains to memory systems with identical scrubbing intervals, where the reciprocal of the mean time between faults equals the scrubbing interval (1 / $\mu$ = T). Summarizing these findings, Table~\ref{tab5} presents the MTTF for memory systems employing deterministic scrubbing and confirms the mentioned conclusions.

The MTTF of the memory system with mixed scrubbing is relatively less influenced by changes in the values of a single parameter, T or $\mu$. This is attributed to the combined effect of deterministic and probabilistic scrubbing applied to the memory words. This distinct behaviour is evident when compared to memory systems employing either probabilistic or deterministic scrubbing alone, as illustrated in Table~\ref{tab6}.

In Table~\ref{tab6}-a, it is evident that for $\mu=0.01$, the MTTF of the memory system utilizing probabilistic scrubbing is 130 days. In contrast, the MTTF for the system employing mixed scrubbing is significantly higher at 2618 days. This showcases that with a tenfold increase in the parameter $1/\mu$ (from 10 to 100), the MTTF of the first system reduces by tenfold. In contrast, the MTTF of the memory system with mixed scrubbing only experiences a 24\% decrease. Further reducing the parameter $\mu$ by tenfold ($\mu=0.001$), the MTTF of the first system reduces by tenfold, whereas the MTTF of the memory system with mixed scrubbing only decreases by 3\%.

Table~\ref{tab6}-b supports similar conclusions, indicating that a tenfold increase in the parameter T (from 10 to 100) results in a 58\% decrease in the MTTF of the memory system with deterministic scrubbing.

Lastly, Table~\ref{tab6}-c demonstrates that the MTTF of memory systems employing mixed scrubbing is affected similarly to systems with either deterministic or probabilistic scrubbing when both parameters T and $1/\mu$ are altered.

Fig.~\ref{fig:4.3} and Fig.~\ref{fig:4.4}  illustrate the reliability of memory systems concerning error rates and memory size. These figures clearly demonstrate that mixed scrubbing yields the highest reliability, while probabilistic scrubbing results in the lowest reliability when T=1/$\mu$.

In Figure~\ref{fig:4.7}, we can observe the reliability of the three memory systems for various values of T and 1/$\mu$. The systems being compared in these figures are configured with 32-bit words, a memory size of 128 MB, and an error rate of 0.00001.

\begin{figure}[htbp]
\centering
\includegraphics[width=3.3in]{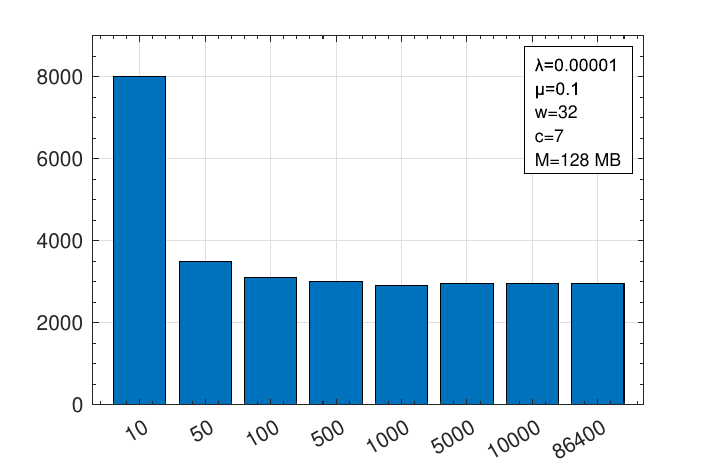}
\caption{The MTTF of mixed scrubbing for various T.}
\label{fig:5.1}
\end{figure}

\begin{table}[htbp]
\caption{The MTTF of the Proposed Mixed Scrubbing Compared with Simple Deterministic or Probabilistic Scrubbing} 
\label{tab5.1}
\centering
\begin{tabular}{ccc}
 & MTTF (days)& Complexity \\
\hline\hline
Probabilistic (1/$\mu$=10) & 2811 & Simple\\  \\
Deterministic (T=20 sec) & 2836 & Very High\\ \\
Mixed (T=1 day) & 2896 & Simple\\ 
\hline
\end{tabular}
\end{table}

Figure~\ref{fig:5.1} illustrates the variation in the MTTF of the memory system utilizing mixed scrubbing for different values of the parameter T. It is evident from Figure~\ref{fig:5.1} that the MTTF of the mixed system experiences a gradual decrease before eventually stabilizing at a value closely approximating that of the system employing probabilistic scrubbing, which is 2811 days.
Considering the information presented above, we recommend the adoption of mixed scrubbing, particularly when deterministic scrubbing is performed on a daily basis. This adjustment results in reduced hardware complexity, as the T interval can be relatively large. The achieved MTTF is superior to that of probabilistic scrubbing with parameters $\mu$=0.1 and deterministic scrubbing with a parameter T=20. The corresponding results are detailed in Table~\ref{tab5.1}.

\section{Conclusions}

In this paper, we have explored and analyzed the reliability of fault-tolerant memory systems using three distinct scrubbing techniques: exponentially distributed scrubbing, deterministic scrubbing, and mixed scrubbing. These models employed Single-Error Correction and Double-Error Detection (SEC-DED) codes and aimed to ensure system reliability in the face of environmental disruptions and inherent circuit weaknesses. Our findings reveal that mixed scrubbing outperforms probabilistic and deterministic scrubbing methods in terms of reliability and MTTF. The mixed scrubbing model's ability to combine the strengths of probabilistic and deterministic approaches while maintaining a simpler architecture, particularly when utilizing a relatively large scrubbing interval, stands out as a key advantage. This approach not only enhances system reliability but also extends the MTTF, making it a compelling choice for memory systems.

The findings of this research suggest that mixed scrubbing is a promising approach for improving the reliability of fault-tolerant memory systems, particularly in scenarios where daily deterministic scrubbing is performed. It offers a robust solution for addressing transient errors and environmental disruptions, making it a valuable addition to the design of fault-tolerant memory systems.

\balance
\bibliographystyle{IEEEtran}
\bibliography{SEEDA}

\end{document}